\input{aipcheck}

\documentclass[
    ,final            
  ]
  {aipproc}

\layoutstyle{6x9}

\begin{document}

\title{A resonance interpretation for the nonmonotonic behavior of the $\phi$ photoproduction cross section near threshold}


\author{Shin Nan Yang}{
  address={Department of Physics, National Taiwan University,
Taipei 10617, Taiwan}
  ,altaddress={Center for Theoretical Sciences, National Taiwan University,
Taipei 10617, Taiwan} 
}

\author{Alvin Kiswandhi}{
  address={Department of Physics, National Taiwan University,
Taipei 10617, Taiwan}
  ,altaddress={Center for Theoretical Sciences, National Taiwan University,
Taipei 10617, Taiwan}
}

\author{Ju-Jun Xie}{
  address={Department of Physics, National Taiwan University,
Taipei 10617, Taiwan}
}

\begin{abstract}
We study whether the nonmonotonic behavior found in  the
differential cross section of the $\phi$-meson photoproduction near
threshold can be described by a resonance. The resonant
contribution is evaluated by using an effective Lagrangian approach. We find
that, with the assumption of a $J^P=3/2^-$ resonance with  mass of
$2.10\pm 0.03$ GeV and width of $ 0.465\pm 0.141$ GeV, LEPS data can
indeed be well described. The ratio of the helicity amplitudes
$A_{\frac 12}/A_{\frac 32}$ calculated from the resulting coupling
constants differs in sign from that of the known $D_{13}(2080)$. We
further find that the addition of this postulated resonance can
substantially improve the agreement between the existing theoretical
predictions and the recent $\omega$ photoproduction data if  a large
value of the OZI evading parameter $x_{\mathrm{OZI}}=12$ is assumed for the
resonance.
\end{abstract}

\maketitle



Recently, a local maximum in the differential cross sections (DCS)
of $\phi$ photoproduction on  protons at  forward angles at around
$E_{\gamma}\sim 2.0$ GeV, has been observed by the LEPS
collaboration \cite{leps05}. Models which consist of $t$-channel
exchanges Refs. \cite{titov97}-\cite{titov07}
have not been able to account  for such a nonmonotonic behavior.

Here, we study whether the nonmonotonic behavior found in
Ref. \cite{leps05} can be described by a resonance. Namely, we will add a
resonance to a model consisting of Pomeron and $(\pi,
\eta)$ exchange \cite{bauer78,donnachie87} by fiat and see if, with a suitable assignment of spin
and parity, mass and width, as well as the coupling constants, one
would be able to obtain a good description of all the data reported
by the LEPS collaboration. Since the local maximum appears quite close to the threshold, we will investigate,
as a first step, the possibility of the spin of the resonance being
either $1/2$ or $3/2$. Similar analysis was carried out in a
coupled-channel model \cite{ozaki09}. However, the analysis was
marred by a confusion in the phase of the Pomeron-exchange amplitude
\cite{hosaka10}.


\begin{figure}[t]

\includegraphics[height=0.21\textheight,angle=0]{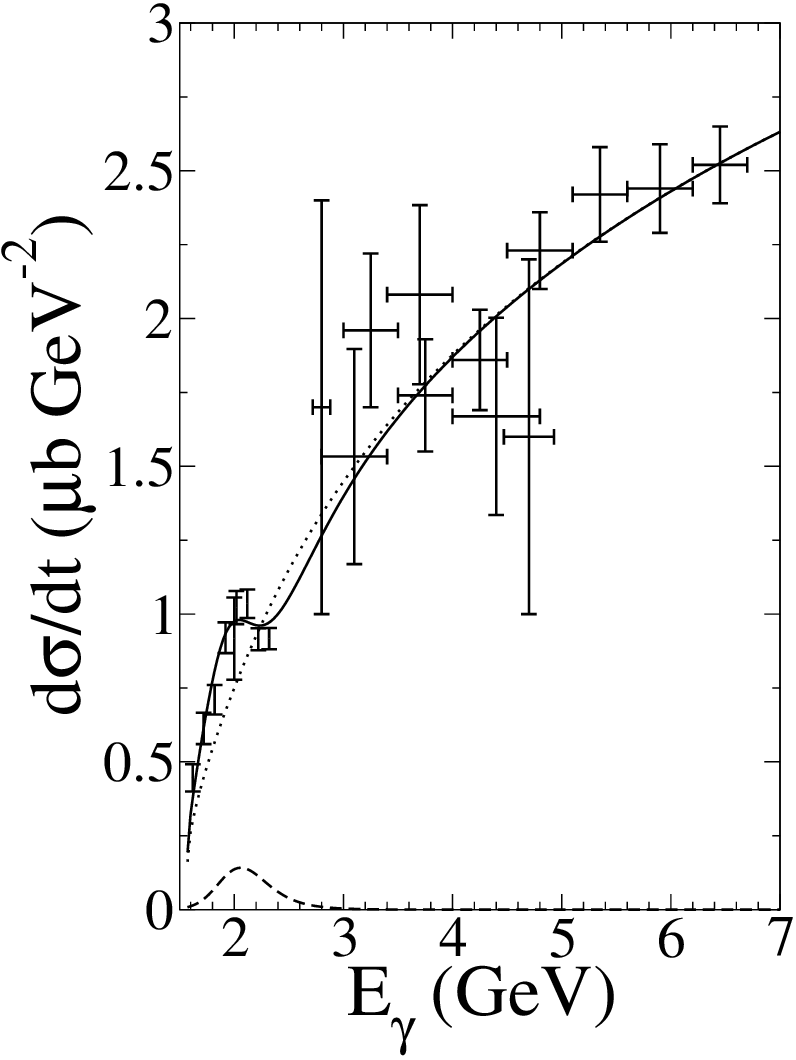}

\hspace{0.3cm}

\includegraphics[height=0.21\textheight,angle=0]{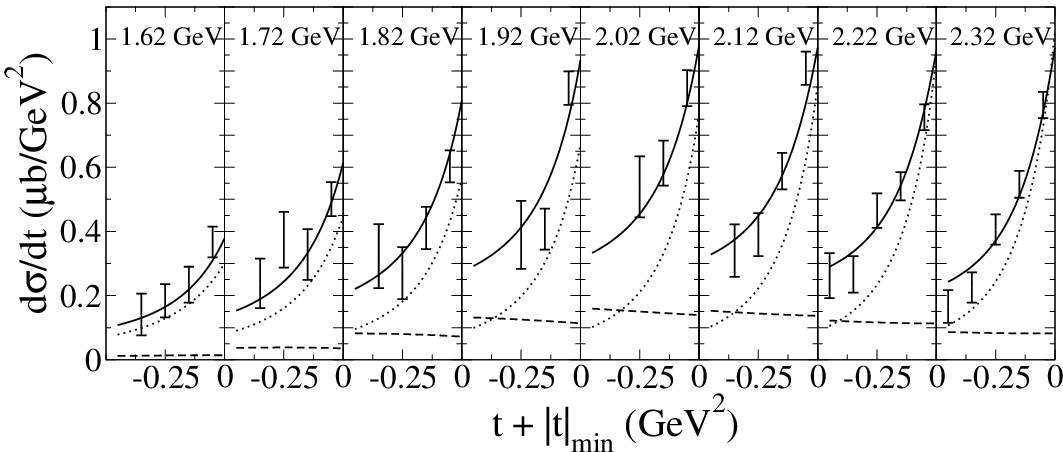}

\caption{Our results for the DCS of $\gamma p \to \phi p$ at
forward direction as a function of photon energy $E_\gamma$ (left) and as a function of $t$
at eight different photon LAB energies (right). Data are from Refs.~\cite{leps05,durham}. The
dotted, dashed, and solid lines denote contributions from nonresonant, resonance with $J^P = 3/2^-$, and their sum,
respectively.
\vspace{-0.75cm} }
\label{D13_1}
\end{figure}

In tree-level approximation, only the mass, width, and the products of coupling constants enter in
the invariant amplitudes. The details of the amplitudes
can be found in Ref. \cite{kiswandhi10}. They are determined with the use of MINUIT,
by fitting to the LEPS experimental data \cite{leps05}.

We found that it is not possible to describe the
nonmonotonic behavior of the DCS at forward direction as a function
of photon energy with only the nonresonant contribution. Furthermore, with an addition of a $J^P = 1/2^\pm$ resonance also cannot produce a nonmonotonic behavior
near threshold, in contrast to the finding of Refs.
\cite{ozaki09,hosaka10}. 

We found that both $J^P = 3/2^\pm$ resonances can describe the data reasonably well.
However, the extracted properties of the $J^P = 3/2^-$
resonance are more stable against changes in Pomeron parameters compared to that of $J^P = 3/2^+$, 
hence our preference of $J^P = 3/2^-$ with mass and width of $2.10 \pm 0.03$ and $0.465 \pm  0.141$, respectively.
The resulting coupling constants are 
$eg_{\gamma N N^*}^{(1)}g_{\phi N N^*}^{(1)} = -0.186 \pm  0.079$, 
$eg_{\gamma N N^*}^{(1)}g_{\phi N N^*}^{(2)} = -0.015 \pm  0.030$, 
$eg_{\gamma N N^*}^{(1)}g_{\phi N N^*}^{(3)} = -0.02 \pm  0.032$, 
$eg_{\gamma N N^*}^{(2)}g_{\phi N N^*}^{(1)} = -0.212 \pm  0.076$, 
$eg_{\gamma N N^*}^{(2)}g_{\phi N N^*}^{(2)} = -0.017 \pm  0.035$, 
and $eg_{\gamma N N^*}^{(2)}g_{\phi N N^*}^{(3)} = -0.025 \pm  0.037$ \cite{kiswandhi10}.




Our best fits with $J^P = 3/2^-$ to the experimental
energy dependence of the DCS at forward angle and angular dependence of the DCS
\cite{leps05,durham} are shown in Fig.~\ref{D13_1}. 
One sees from 
Fig.~\ref{D13_1} that the resonance improves the agreement with the data

\vspace{0.7cm}

\begin{figure}[htbp]

\includegraphics[height=0.25\textheight,angle=0]{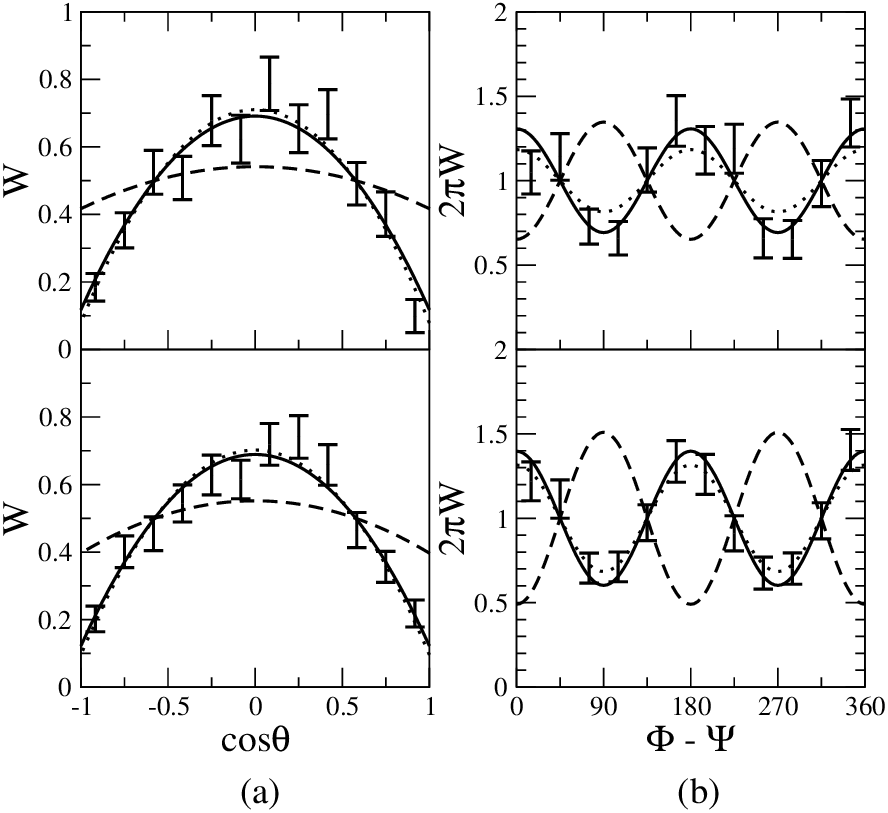}

\hspace{0.5cm}

\includegraphics[height=0.25\textheight,angle=0]{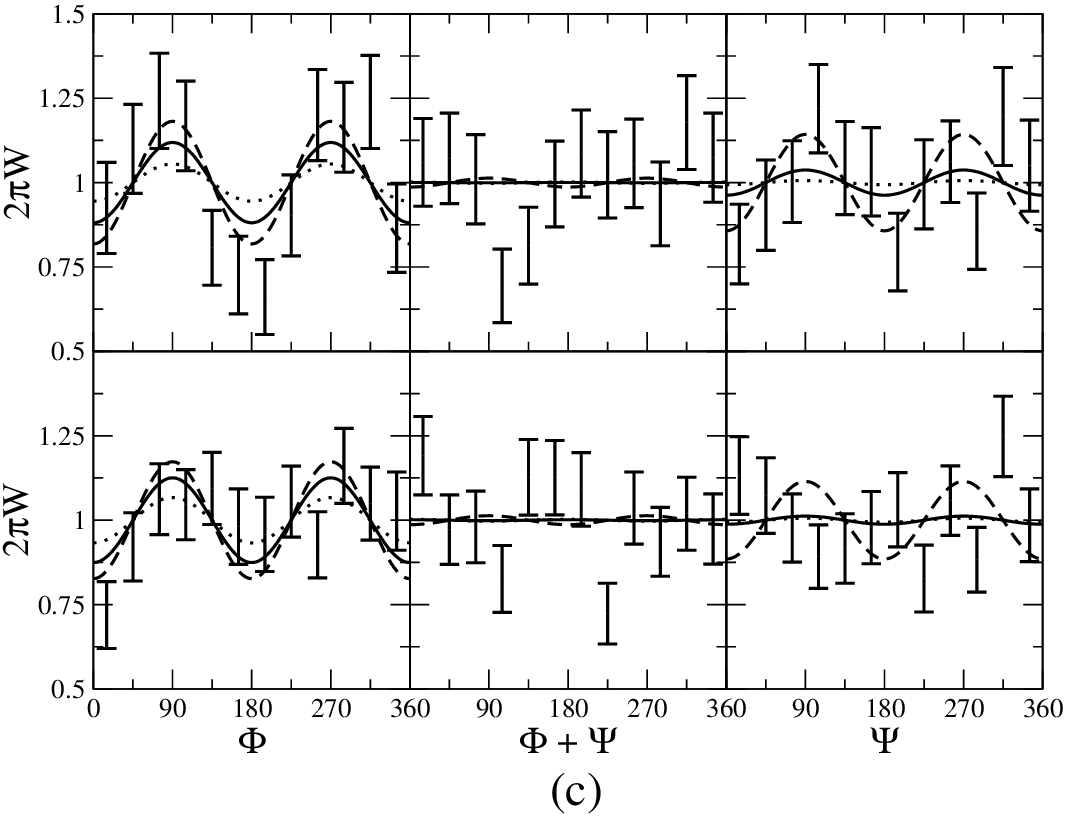}

\caption{{Our results obtained with  $J^P = 3/2^-$ resonance:
(a) decay angular distributions $ W(\cos \theta)$
(b) $W(\Phi - \Psi)$, and (c) $W(\Phi)$,
 $W(\Phi + \Psi)$, and $W(\Psi)$. All the decay angular
 distributions are given in two photon LAB energies,
 $1.97 - 2.17$ GeV (upper panel) and $2.17 - 2.37$ GeV (lower panel).
 Data is taken from Ref.~\cite{leps05}. The notation is the same as in Fig.~\ref{D13_1}. \vspace{-0.cm}}}
 \label{D13_3}
\end{figure}

Our results for the decay angular distributions of the $\phi$-meson
in the Gottfried-Jackson system (hereafter,
called GJ-frame)~\cite{titov03,schillingnpb15397}, are shown in
Fig. \ref{D13_3}. Here, the inclusion of resonant contribution does help the agreement with the data, 
especially for $W(\Phi-\Psi)$  at $2.17-2.37$ GeV and $W(\Phi)$ at $1.97-2.17$ GeV.
For both $W(\Phi+\Psi)$ and $W(\Psi)$, our model
still fail to give adequate agreement with the data which are of
rather poor quality with large error bars.

One might be tempted to identify the $3/2^-$ as the $D_{13}(2080)$
as listed in PDG \cite{PDG02}. However, with the coupling constants given above, we obtain a value of
{$A_{1 \over 2}/A_{3 \over 2}=1.16$} which differ from
$-1.18$ for $D_{13}(2080)$ \cite{PDG02} in relative sign.

\begin{figure}[htbp]

\includegraphics[height=0.25\textheight,angle=0]{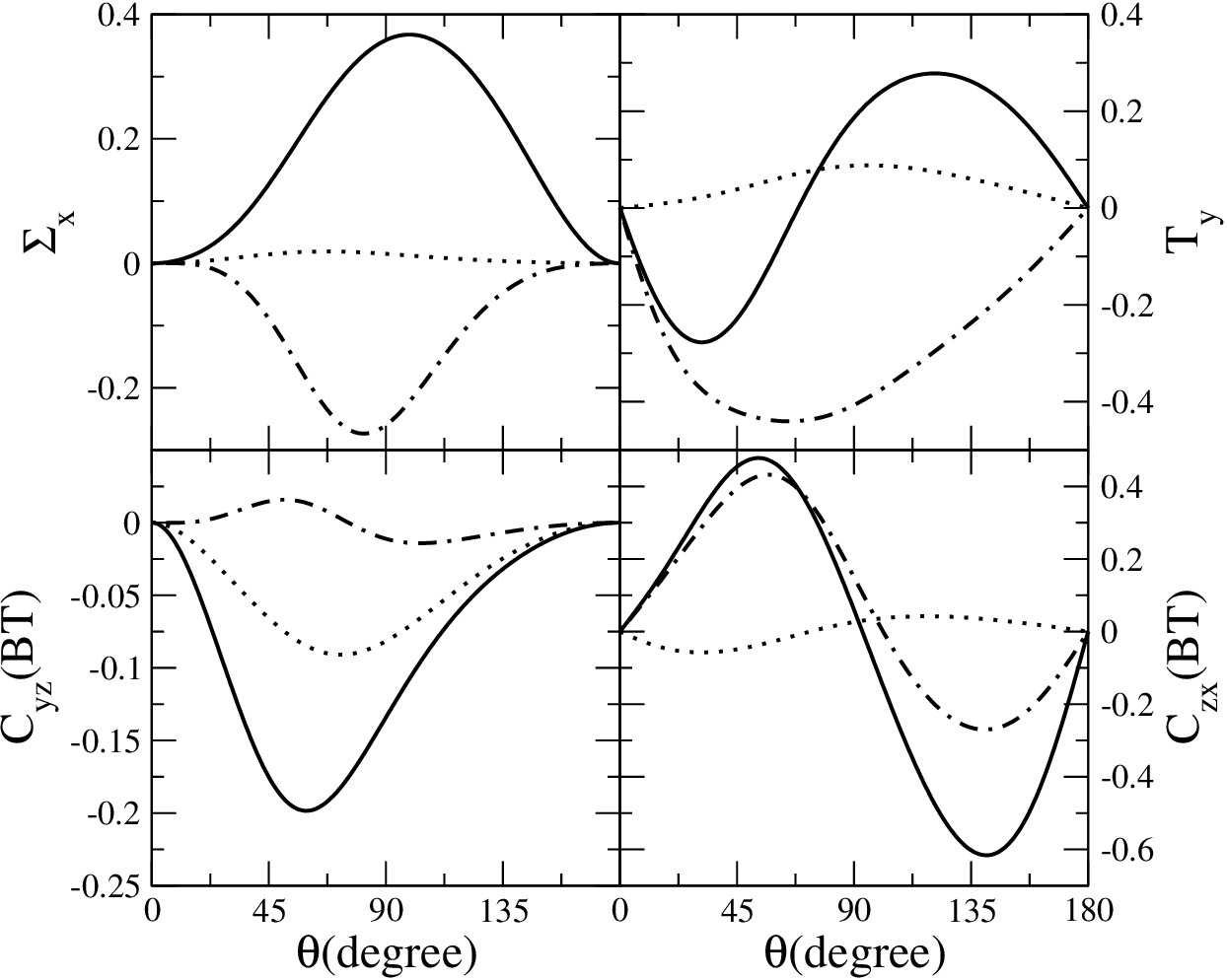}

\includegraphics[height=0.25\textheight,angle=0]{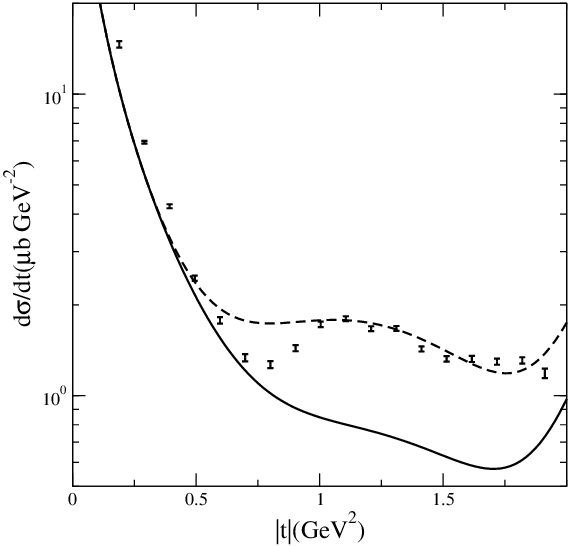}

\caption{{Left: Single and double polarization observables
$\Sigma_x$, $T_y$, $C_{yz}^{BT}$, and $C_{zx}^{BT}$ taken at photon
laboratory energy $E_\gamma = 2$ GeV. The solid and dash-dotted
lines correspond to our results with the choices of $J^P = 3/2^-$ and
$J^P = 3/2^+$, respectively, while the dotted lines denote the
nonresonant contribution. Right: DCS of $\omega$ photoproduction as a
function of $|t|$ at $W = 2.105$ GeV. Solid and dashed lines
represent the model predictions of Ref. \cite{oh02} without and with
the addition of our resonance   with $x_{\mathrm{OZI}} = 12$. Data are
from Ref. \cite{M_Williams}.}} \label{Polarization}

\end{figure}

In general, we find that the effects of the
resonance are substantial in many of the polarization observables {\cite{titov_polarization}}. Results for 
single and double polarization observables $\Sigma_x$, $T_y$, $C_{yz}^{BT}$, and $C_{zx}^{BT}$ are shown in left panel of Fig.~\ref{Polarization}. 
In the same figure, the results using a $3/2^+$ resonance are also shown by dash-dotted curve. We see that
measurements of these polarization observables would help to
resolve the question of the parity of the resonance.


\vspace{0.785cm}

From the $\phi-\omega$ mixing, one would expect that a resonance
in $\phi N$ channel would  also appear in $\omega N$ channel.
The conventional "minimal" parametrization relating $\phi NN^*$ and
 $\omega NN^*$ is $g_{\phi N N^*} = -\tan \Delta \theta_V
x_{\mathrm{OZI}} g_{\omega N N^*}$, with $\Delta \theta_V \simeq 3.7^\circ$ corresponds to the
deviation from the ideal $\phi-\omega$ mixing angle.
The larger value of the OZI-evading parameter $x_{\mathrm{OZI}}$ would indicate larger strangeness
content of the  resonance.

By adding the resonance
postulated here to the model of Ref. \cite{oh02} with $
x_{\mathrm{OZI}}=12$, whose prediction is given in the dashed line in right panel of 
Fig.~\ref{Polarization}, we see that the DCS at $W = 2.105$ GeV can be
reproduced with roughly the correct strength. The large value of
$x_{\mathrm{OZI}}=12$ would imply that the resonance we propose here
contains a considerable amount of strangeness content.

In summary, we study the possibility of accounting for the
nonmonotonic behavior as observed by the LEPS collaboration at
energies close to threshold as a possible manifestation of a resonance.
We confirm that nonresonant contribution alone cannot describe the
LEPS data, as well as the addition of a resonance with $J = 1/2$. However, with
an assignment of $J=3/2^-$, a nice agreement with most of the LEPS
data can be achieved, with a greater stability  
with respect to changes in the low-energy Pomeron parameters, compared to $J=3/2^+$.
The obtained resonance mass and width are
$2.10\pm 0.03$ and $ 0.465\pm 0.141$ GeV, respectively.  The
resulting coupling constants give rise to a ratio of the helicity
amplitudes that differs from that of the known $D_{13}(2080)$ in sign.
Furthermore, we find that the postulated resonance gives substantial
contribution to the polarization observables, which can also be used
to determine the parity of the resonance if it indeed exists.
The addition of a $J=3/2^-$ resonance to the model of Ref. \cite{oh02} with a choice of a large value of OZI-evading
parameter $x_{\mathrm{OZI}}=12$ could indeed considerably improve
the agreement of the model prediction with the most recent data.
That would imply the resonance postulated here does contain
considerable amount of strangeness content.



\begin{theacknowledgments}
We would like to thank Profs. W. C. Chang, C. W. Kao, Atsushi Hosaka,
T.-S. H. Lee, Yongseok Oh, Sho Ozaki, and A. I. Titov, for useful
discussions and/or correspondences. We also acknowledge the help from National Taiwan University High-Performance Computing Center. This work is supported in part
by the National Science Council of R.O.C. (Taiwan) under grant
NSC99-2112-M002-011.
\end{theacknowledgments}



\bibliographystyle{aipproc}   

\bibliography{sample}

\begin{thebibliography}{00}
%
\bibitem{leps05}T. Mibe {\it et al.} (LEPS Collaboration), \emph{Phys. Rev.
Lett.} {\bf 95}, 182201 (2005) and references therein.
%
\bibitem{titov97}A. I. Titov, Y. S. Oh and S. N. Yang, \emph{Phys. Rev. Lett.} {\bf 79}, 1634
(1997); A. I. Titov, Y. Oh, S. N. Yang, and T. Morii, \emph{Nucl. Phys. A} {\bf 684}, 354 (2001).
%
\bibitem{titov_polarization} {A. I. Titov, Y. Oh, S. N. Yang, and T. Morii, \emph{Phys. Rev. C}
{\bf 58}, 2429 (1998).}
%
\bibitem{williams98}R. A. Williams, \emph{Phys. Rev. C }{\bf 57}, 223 (1998).
%
\bibitem{titov99}A. I. Titov, T.-S. H. Lee, H. Toki, and O. Streltsova,
\emph{Phys. Rev. C} {\bf 60}, 035205 (1999).
%
%
\bibitem{titov03}A. I. Titov, T.-S. H. Lee, \emph{Phys. Rev. C} {\bf 67}, 065205
(2003).
%
\bibitem{titov07} A. I. Titov and B. K\"ampfer, \emph{Phys. Rev. C} {\bf 76}, 035202 (2007).
%
\bibitem{bauer78}T. H. Bauer, R. D. Spital, D. R. Yennie and F. M.
Pipkin, \emph{Rev. Mod. Phys.} {\bf 50}, 261 (1978).
%
\bibitem{donnachie87}A. Donnachie and P. V. Landshoff, \emph{Phys. Lett. B} {\bf
185}, 403 (1987); \emph{Nucl. Phys. B} {\bf 244}, 322 (1984); \emph{Nucl. Phys. B}
{\bf 267}, 690 (1986); \emph{Nucl. Phys. B} {\bf 311}, 509 (1989).
%
%
%
%
\bibitem{ozaki09} S. Ozaki, A. Hosaka, H. Nagahiro, and O. Scholten, \emph{Phys. Rev. C} {\bf 80}, 035201 (2009).
%
\bibitem{hosaka10}A. Hosaka, private communication.
%
\bibitem{kiswandhi10} A. Kiswandhi, J.-J. Xie, S.N. Yang, \emph{Phys. Lett. B} {\bf 691}, 214-218 (2010) and references therein.
%
\bibitem{durham} Durham HEP database (http://www.slac.stanford.edu/spires/hepdata).
%
%
%
%
%
%
%
\bibitem{schillingnpb15397}K. Schilling, K. Seyboth, and G. Wolf,
\emph{Nucl. Phys. B} {\bf 15}, 397 (1970).
%
\bibitem{PDG02}Particle Data Group, K. Hagiwara {\it et al.}, \emph{Phys.
Rev. D} {\bf 66}, 010001 (2002).
%
%
%
\bibitem{oh02} Y. Oh, A. I. Titov, and T.-S.H. Lee, \emph{Phys. Rev. C }{\bf 63}, 025201 (2001).
%
\bibitem{M_Williams} M. Williams et.al., \emph{Phys. Rev. C }{\bf 80}, 065208 (2009).
%
\end{thebibliography}

\end{document}